\documentclass{article} 
\usepackage{emulateapj} 
\usepackage{apjfonts}
\usepackage{epsfig}

\begin{document}

\title{Constraints on Cosmological Parameters from MAXIMA-1}

\author {A.~Balbi\altaffilmark{1,2,3}, P.~Ade\altaffilmark{4},
  J.~Bock\altaffilmark{5,6}, J.~Borrill\altaffilmark{7,2},
  A.~Boscaleri\altaffilmark{8}, P.~De Bernardis\altaffilmark{9}
  P.~G.~Ferreira\altaffilmark{10,11}, S.~Hanany\altaffilmark{12,2},
  V.~Hristov\altaffilmark{6}, A.~H.~Jaffe\altaffilmark{2,14},
  A.~T.~Lee\altaffilmark{2,3,15},
S.~Oh\altaffilmark{2,15},
  E.~Pascale\altaffilmark{8}, B.~Rabii\altaffilmark{2,15},
  P.~L.~Richards\altaffilmark{15,2},
  G.~F.~Smoot\altaffilmark{15,2,3,14},
  R.~Stompor\altaffilmark{2,14,17}, C.~D.~Winant\altaffilmark{2,15},
  J.~H.~P.~Wu\altaffilmark{16}}

\altaffiltext{1}{Dipartimento di Fisica, Universit\`a Tor Vergata,
  Roma, Italy}

\altaffiltext{2}{Center for Particle Astrophysics, University of
  California, Berkeley, CA, USA}

\altaffiltext{3}{Lawrence Berkeley National Laboratory, University of
  California, Berkeley, CA, USA}

\altaffiltext{4}{Queen Mary and Westfield College, London, UK}

\altaffiltext{5}{Jet Propulsion Laboratory, Pasadena, CA, USA}

\altaffiltext{6}{California Institute of Technology, Pasadena, CA,
  USA}

\altaffiltext{7}{National Energy Research Scientific Computing Center,
  Lawrence Berkeley National Laboratory, Berkeley, CA, USA}

\altaffiltext{8}{IROE-CNR, Florence, Italy}

\altaffiltext{9}{Dipartimento di Fisica, Universit\`a La Sapienza,
  Roma, Italy}

\altaffiltext{10}{Astrophysics, University of Oxford, UK}

\altaffiltext{11}{CENTRA, Instituto Superior Tecnico, Lisboa,
  Portugal}

\altaffiltext{12}{School of Physics and Astronomy, University of
  Minnesota/Twin Cities, Minneapolis, MN, USA}

\altaffiltext{13}{Dept. of Physics and Astronomy, University of
  Massachussets, Amherst, MA, USA}

\altaffiltext{14}{Space Sciences Laboratory, University of California,
  Berkeley, CA, USA}

\altaffiltext{15}{Dept. of Physics, University of California,
  Berkeley, CA, USA}

\altaffiltext{16}{Dept. of Astronomy, University of California,
  Berkeley, CA, USA}

\altaffiltext{17}{Copernicus Astronomical Center, Warszawa, Poland}

\begin{abstract}
  We set new constraints on a seven-dimensional space of cosmological
  parameters within the class of inflationary adiabatic models. We use
  the angular power spectrum of the cosmic microwave background
  measured over a wide range of $\ell$ in the first flight of the
  MAXIMA~balloon-borne experiment (MAXIMA-1) and the low $\ell$
  results from COBE/DMR.  We find constraints on the total energy
  density of the universe, $\Omega=1.0^{+0.15}_{-0.30} $, the physical
density
  of baryons, $\Omega_{\rm b}h^2=0.03\pm0.01$, the physical density
  of cold dark matter, $\Omega_{\rm cdm}h^2=0.2^{+0.2}_{-0.1}$, and the
  spectral index of primordial scalar fluctuations, $n_s=1.08\pm 0.1$,
  all at the $95\%$ confidence level.  By combining our results
  with measurements of high-redshift supernovae we constrain the value
  of the cosmological constant and the fractional amount of
  pressureless matter in the universe to $0.45<\Omega_\Lambda<0.75$
  and $0.25<\Omega_{\rm m}<0.50$, at the $95\%$ confidence level.  Our
  results are consistent with a flat universe and the shape parameter 
  deduced from large scale structure, and in marginal agreement with the
  baryon density from big bang nucleosynthesis.
\end{abstract}


\section{Introduction}

The angular power spectrum of the cosmic microwave background (CMB)
temperature anisotropy depends strongly on the values of most
cosmological parameters and on the physical processes in the early
universe; it is thus a powerful tool to constrain cosmological models
(see e.g., Kamionkowski \& Kosowsky 1999).  Several authors have used CMB data
from a number of experiments to constrain cosmological parameters (see
e.g., Lange et al. 2000, Tegmark \& Zaldarriaga 2000, 
and references therein).

In this letter we make use of the angular power spectrum measured in
the first flight of the MAXIMA~balloon-borne experiment, MAXIMA-1
(described in a companion paper by Hanany et al. 2000).  This power
spectrum covers a range $36\leq \ell \leq 785$, which is the largest
coverage in multipole space from a single experiment to date.  We
include in our analysis the 4-year COBE/DMR angular power spectrum
(G\'orski et al. 1996), which spans the range $2\leq \ell \leq 34$, to
normalize the models at large angular scales.  We determine
constraints on a seven-dimensional space of cosmological parameters
within the class of inflationary adiabatic models.

This letter is structured as follows. In Section \ref{methods} we
describe and justify the cosmological parameters space we have
explored and the technique we used to perform the maximum likelihood
analysis.  In Section \ref{constraints} we apply the method to the
MAXIMA-1 and COBE/DMR data, to set constraints on the suite of
parameters. We also combine our constraints with those from the
high-redshift supernovae measurements (Perlmutter et al. 1999; Riess
et al. 1998) to set bounds on the fractional density in a cosmological
constant, $\Omega_\Lambda$, and pressureless matter, $\Omega_{\rm m}$.
In Section \ref{discussion} we summarize our results and discuss
their implications for fundamental models of structure formation.

\section{Dataset, Methods and Parameter Space}
\label{methods}

The MAXIMA-1 power spectrum is estimated in 10 bins spanning the range
$36\leq \ell \leq 785$ (Hanany et al. 2000).  In each bin, the spectrum is
assigned a flat shape, $\ell(\ell+1)C_\ell/2\pi={\cal C}_B$, whose
amplitude is found by maximizing the likelihood of the data using a
quadratic estimator technique (Bond, Jaffe \& Knox 1998) as
implemented in the MADCAP software package (Borrill 1999).  We find
the most likely values of the cosmological parameters by maximizing
the likelihood of the data.  Following Bond, Jaffe \& Knox (2000) we use
a likelihood which is Gaussian in the quantity $\ln({\cal C}_B+x_B)$,
with $x_B$ related to the noise properties of the experiment.

In this letter we only consider inflationary adiabatic models.  There
are two main reasons.  First, the poor performance of alternative
theories in reconciling measurements of clustering on galactic scales
with measurements of CMB fluctuations on horizon scales has made
inflation the favored contender as a fundamental theory of structure
formation.  Second, the angular power spectra for inflationary models
are easy to calculate for given cosmological parameters, especially
since the advent of fast Einstein-Boltzmann solvers such as CMBFAST
(Seljak \& Zaldarriaga 1996) or CAMB (Lewis, Challinor \& Lasenby
2000).

We consider a seven-dimensional space of parameters. This includes the
amplitude of fluctuations at $\ell=10$, $C_{10}$, the fractional
densities of baryons, $\Omega_{\rm b}$, 
and cosmological constant, $\Omega_\Lambda$, 
the total energy density of the universe, 
$\Omega\equiv \Omega_{\rm m}+\Omega_\Lambda$ (where the fractional
density of pressureless matter is
$\Omega_{\rm m}\equiv \Omega_{\rm b}+\Omega_{\rm cdm}$ and 
$\Omega_{\rm cdm}$ is the fractional density of cold dark matter)  
the spectral index of primordial scalar fluctuations, $n_s$, and the
optical depth of reionization, $\tau_c$. We use the following ranges
and sampling: $C_{10}$ is continuous; $\Omega$ = 0.1, 0.3, 0.5, 0.7, 0.8, 
0.85, 0.9, 0.95, 1, 1.05, 1.1, 1.15, 1.2, 1.3, 1.5;
$\Omega_{\rm b}$ = 0.01, 0.03, 0.05, 0.07, 0.09,
0.12, 0.20, 0.28; $\Omega_\Lambda$ = 0.0, 0.1, $\cdots$, 0.8, 0.9;
$n_s$ = 0.60, 0.64, $\cdots$, 1.36, 1.40; 
$\tau_c$ = 0, 0.025, 0.05, 0.075, 0.1, 0.15, 0.2, 0.3, 0.5.  For
the seventh parameter we have either used the Hubble parameter,
$H_0$ = 40, 45, $\cdots$, 85, 90 km s$^{-1}$Mpc$^{-1}$, or the physical
baryon density, $\Omega_{\rm b}h^2$ = 0.003125, 00625, 0.0125, 0.0175, 0.02,
0.025, 0.030, 0.035, 0.04, 0.05, 0.075, 0.1, 0.15, 0.2
(where $h\equiv H_0/(100~$km s$^{-1}$Mpc$^{-1})$).  This
approach allows us to check the importance of the parameters ranges,
which are effectively used as a uniform prior in our analysis.  When
we use $H_0$ as the free parameter we get relatively fine but
restricted coverage of $H_0$ and $\Omega_{\rm b}h^2$. When we use
$\Omega_{\rm b}h^2$ we consider a much more extended set of priors.
To ensure that we are sampling the set of parameters with enough
resolution, we include models corresponding to values of the
parameters that are not on the grid by quadratically interpolating
between power spectra.  We have not allowed for the presence of
massive neutrinos, since their effect on the
angular power spectrum of the CMB would be negligible for the dataset
used here (Dodelson, Gates \& Stebbins 1996).

We evaluate the likelihood for a subset of parameters by maximizing it
over all the remaining parameters.  This approach, already adopted in
previous analyses of this kind (Tegmark \& Zaldarriaga 2000;
Melchiorri et al. 1999), replaces the time-consuming integration that
is required to marginalize over the unwanted parameters. In all our
estimates we marginalize over the overall MAXIMA-1 calibration
uncertainty (Hanany et al. 2000), assuming a Gaussian prior; we have
also marginalized over the beam uncertainty reported in
Hanany et al. (2000) but have found that it has a negligible effect on our
results.

\section{Constraints}
\label{constraints}
The position of the first peak in the angular power spectrum of
adiabatic models can be used to constrain the geometry of the universe
(Doroshkevich, Zel'dovich \& Sunyaev 1978).  Features in the
radiation pattern at recombination are set by the maximum distance
sound waves have traveled at that time.  The diameter-distance
relation allows us to relate the physical scale of these features to
the angle they subtend on the sky and strongly depends on the
curvature of the universe, which is determined by the total energy
density of the universe $\Omega\equiv\Omega_{\rm m}+\Omega_\Lambda$.
Other cosmological parameters have much less effect on this relation.
\vskip 0.3cm
\vbox{\epsfxsize=8.5cm\epsfbox{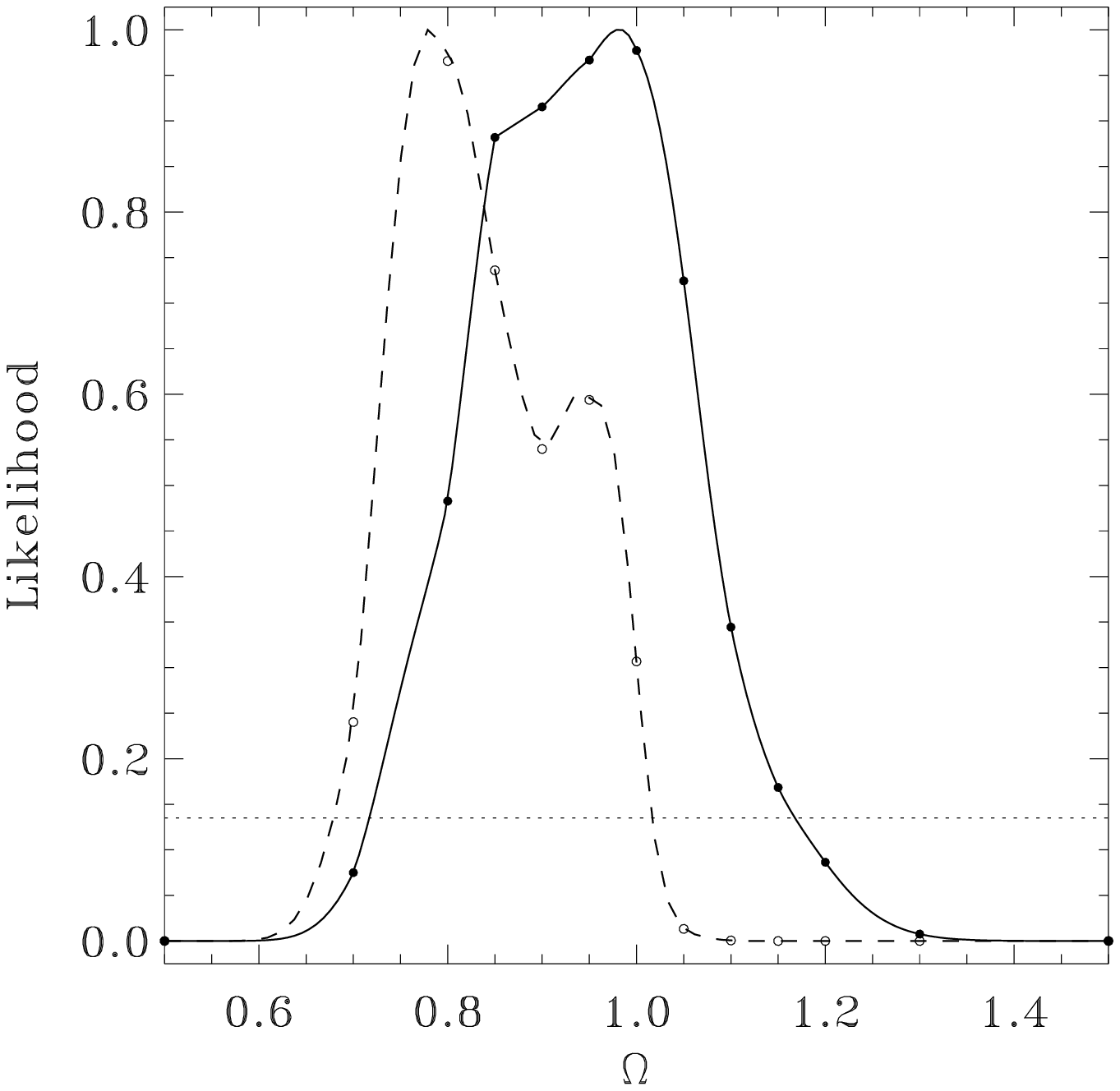}} \vskip 0.3cm { \small
  F{\scriptsize IG}.~1.--- Likelihood function of the total energy 
  density of the universe $\Omega$.  
  The solid line was obtained by maximizing over all other
  parameters over the ranges described in the text, while the dashed
  line was obtained by constraining $\Omega_{\rm b}h^2=0.0190\pm
  0.0024$ and $H_0=65\pm 7$~km~s$^{-1}$~Mpc$^{-1}$.  The intersections
  with the horizontal line give the bounds for $95\%$ confidence.
\label{fig:fig1}}
\vskip 1cm

Figure~1 shows the likelihood of the total energy density of the
universe, $\Omega$.  We obtain the solid line if we maximize over all
the remaining parameters; we get the dashed line if we impose the big
bang nucleosynthesis (BBN) constraint $\Omega_{\rm b}h^2=0.0190\pm
0.0024$ (Burles et al. 1999; see also, Tytler et al. 2000), 
and restrict $H_0=65\pm 7~$km~s$^{-1}$~Mpc$^{-1}$ (Freedman 1999).  
From the solid line we see that $0.7<\Omega<1.15$ at
the $95\%$ confidence level.

We identified the other tightly constrained parameters from a principal 
component analysis of the likelihood function (Efstathiou \& Bond 1999). 
In Figure~2 we plot the
likelihoods for three well-constrained parameters, the physical baryon
density, $\Omega_{\rm b}h^2$, the physical cold dark matter density,
$\Omega_{\rm cdm}h^2$, and the spectral index of primordial scalar
fluctuations, $n_s$. The likelihood for each parameter was obtained by
maximizing over all the remaining parameters.

We find $0.02<\Omega_{\rm b}h^2<0.04$, $0.1 < \Omega_{\rm cdm}h^2
<0.4$, and $0.98<n_s<1.18$, all at the $95\%$ confidence level. The
constraint on $\Omega_{\rm b}h^{2}$ is independent of the one coming
from BBN (Burles et al. 1999) and agrees with it
at about the 2$\sigma$ level. 
However, the most likely value of 
$\Omega_{\rm b}h^{2}$ emerging from our dataset is noticeably
higher than the one from BBN. Imposing the
BBN prior to our dataset has the effect of moving the best fit region
towards open models (see Figure~1). The
constraint on $n_s$ indicates that the combined MAXIMA-1 and COBE/DMR data
are consistent with a scale invariant spectrum, 
with a smaller variance than that estimated using previous datasets 
(see e.g., Lange et al. 2000, Tegmark \& Zaldarriaga 2000).
This result should however be interpreted with some caution,
because it is sensitive to assumptions on the treatment of 
calibration uncertainties and depends in part on the absence of 
tensor modes in our analysis.
\vbox{\vskip 0.5cm \epsfxsize=8.5cm\epsfbox{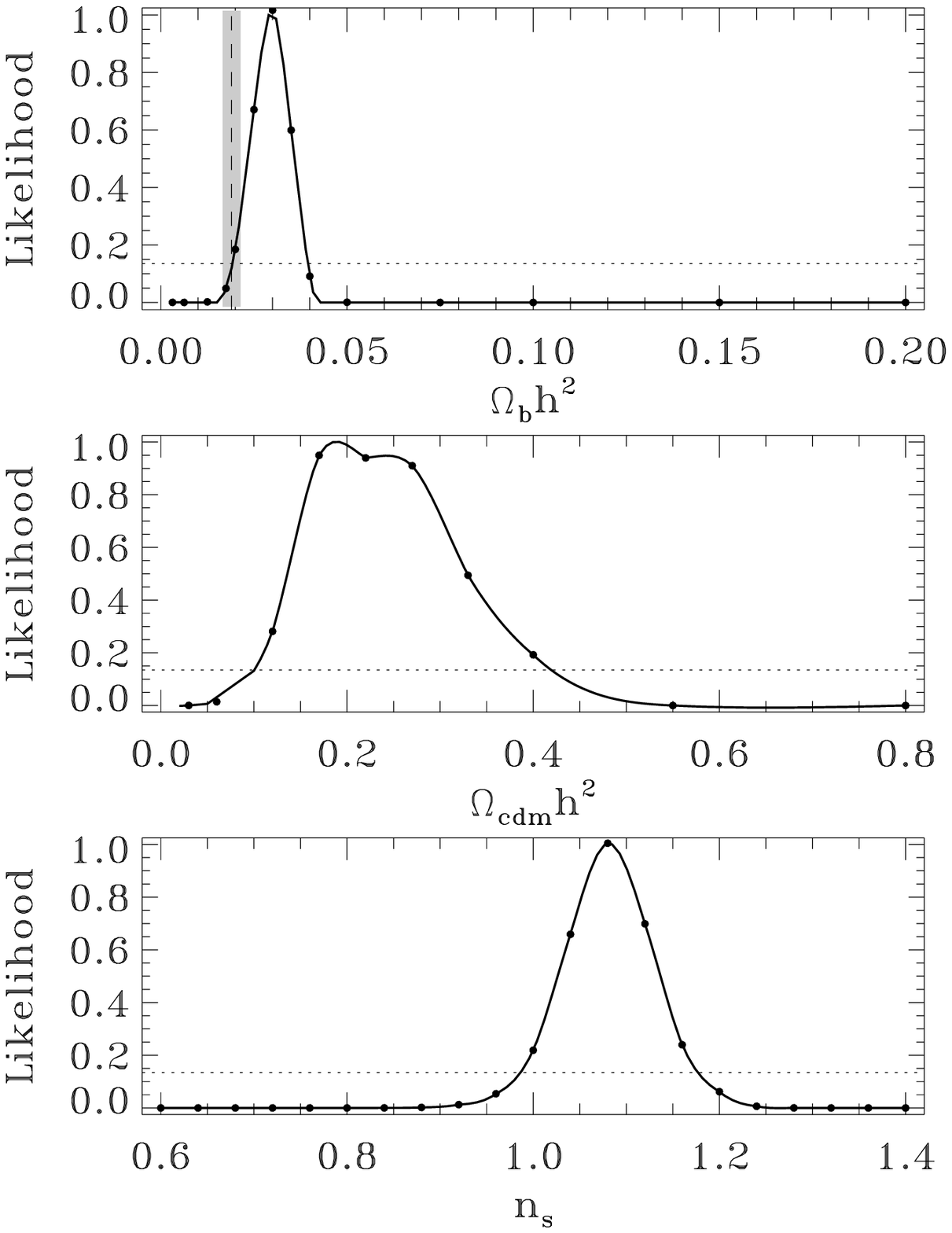}} \vskip 0.5cm
{\small F{\scriptsize IG}.~2.--- Likelihoods of three cosmological parameters: 
  from top to bottom, the physical baryon density, $\Omega_{\rm
    b}h^2$, the physical cold dark matter density, $\Omega_{\rm
    cdm}h^2$, and the scalar spectral index, $n_s$.  For each of these
  parameters, the likelihood was maximized over all the
  remaining parameters.  The vertical band in the top panel represents
  the BBN constraint $\Omega_{\rm b}h^2=0.0190\pm 0.0024$ 
  (Burles et al. 1999). The intersections with the horizontal line give 
  the bounds for $95\%$ confidence.
\label{fig:fig2}} 
\vskip 1cm

One of the intriguing developments arising in contemporary
observational cosmology is the possible existence of a smooth,
negative-pressure component, for example a cosmological constant,
driving an accelerated expansion of the universe (Perlmutter et al.
1999; Riess et al. 1998).  While CMB data are quite powerful in
constraining the total energy density of the universe,
a precise independent determination of $\Omega_{\rm m}$ and
$\Omega_\Lambda$ is limited by the presence of the geometrical degeneracy
discussed in Efstathiou \& Bond (1999) (which can be broken if
a gravitational lensing imprint on the CMB is detected, 
see e.g., Stompor \& Efstathiou 1999). 
We can however find the locus of models in the
$\Omega_{\rm m}$---$\Omega_\Lambda$ plane which are favored by the
combined MAXIMA-1 and COBE/DMR data.  We do so in Figure~3, where once
again we maximize over all the remaining parameters.
The fact that the MAXIMA-1 dataset extends to large values of $\ell$ helps to
narrow the contours along the degeneracy direction $\Omega_{\rm
  m}+\Omega_\Lambda=$~constant. Although a large portion of the
$\Omega_{\rm m}$---$\Omega_\Lambda$ plane is ruled out, one clearly
needs additional data to strongly break the degeneracy.
We then overlay
our results on the likelihood contours from the high-redshift
supernovae data of Perlmutter et al. (1999) and Riess et al. (1998) and
calculate the combined likelihood. The resulting constraints are:
$0.45<\Omega_\Lambda<0.75$ and $0.25<\Omega_{\rm m}<0.50$, at the
$95\%$ confidence level.
\vbox{\hskip -0.6cm
  \epsfxsize=9.0cm\epsfbox{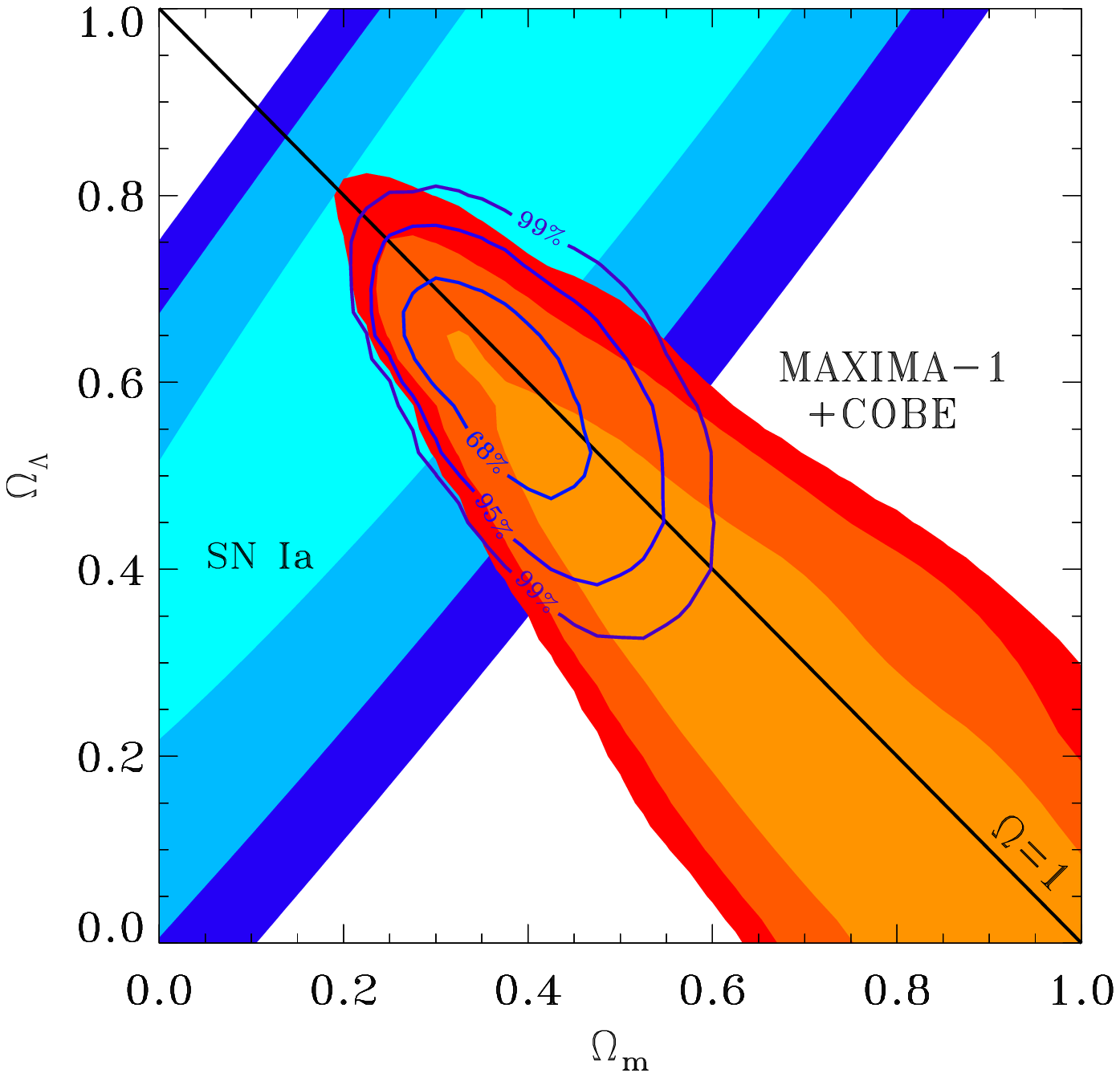}} {\small F{\scriptsize IG}.~3.---
  Constraints in the $\Omega_{\rm m}$--$\Omega_\Lambda$ plane from the
  combined MAXIMA-1 and COBE/DMR datasets. The likelihood at each
  point in the plane is taken by maximizing over the remaining five
  parameters.  The borders of the shaded regions correspond to 0.32,
  0.05 and 0.01 of the peak value of the likelihood.  These are overlaid on the
  bounds obtained from high redshift supernovae data (Perlmutter et
  al. 1999; Riess et al. 1998).  The closed contours are the
  confidence levels from the combined likelihood. The diagonal
  line corresponds to flat models.\label{fig:fig3}} 
\vskip 1cm

\section{Conclusions}
\label{discussion}
Using the broad $\ell$-space coverage of the MAXIMA-1 and COBE/DMR
data we set constraints on a number of cosmological parameters within
the class of inflationary adiabatic models. We find:
$\Omega=1.0^{+0.15}_{-0.30}$, $\Omega_{\rm b}h^2=0.03\pm0.01$, $\Omega_{\rm
  cdm}h^2=0.2^{+0.2}_{-0.1}$ and $n_s=1.08\pm 0.1$ at the $95\%$
confidence level. The constraints we have obtained on $\Omega$ are
consistent with those of other recent analyses (Melchiorri et al.~1999; 
Dodelson \& Knox~1999; Tegmark \& Zaldarriaga~2000; 
Lange et al.~2000) and provide further support to a flat universe.  
Combining our constraints with
those coming from measurements of supernovae at high redshift strongly
favors a non-zero cosmological constant, $\Omega_\Lambda=0.60 \pm
0.15$, and a pressureless matter density parameter $\Omega_{\rm
  m}=0.375 \pm 0.125$, both at the $95\%$ confidence level.

Our analysis was restricted to inflationary adiabatic models.  Several
hints suggest that this may indeed be the right paradigm.  The
presence of a localized peak in our data in the region $150<\ell<250$
is in itself an evidence in favor of
inflationary adiabatic models; alternative theories either predict a
broader peak at higher $\ell$ or a broad shelf at $\ell<200$ (see
e.g., Knox \& Page 1999).  Furthermore, from a goodness-of-fit
analysis (with the caveat that our data are not perfectly 
well-described by a Gaussian likelihood)
we find that the class of models we have considered are
perfectly consistent with the data. Our best fit has $\chi^2$=41
using the 38 data points of the combined MAXIMA-1/COBE dataset 
($\chi^2$=7 using only the 10 data points from MAXIMA-1). Within
inflationary models, the Standard CDM model and the Open CDM model
with $\Omega=0.3$ both provide a poor fit to our data, having,
respectively, $\chi^2$=67 (25) and $\chi^2$=224 (159).  
The $\Lambda$-CDM `concordance model' (Ostriker \& Steinhardt 1995; 
Krauss \& Turner 1995) 
has $\chi^2$=50 (9).
These results are summarized in Figure~4, where we plot the MAXIMA-1 
data overlaid with various cosmological models.
\vskip 0.5cm
\vbox{\vskip -0.8cm \hskip -0.5cm\epsfxsize=9.cm\epsfbox{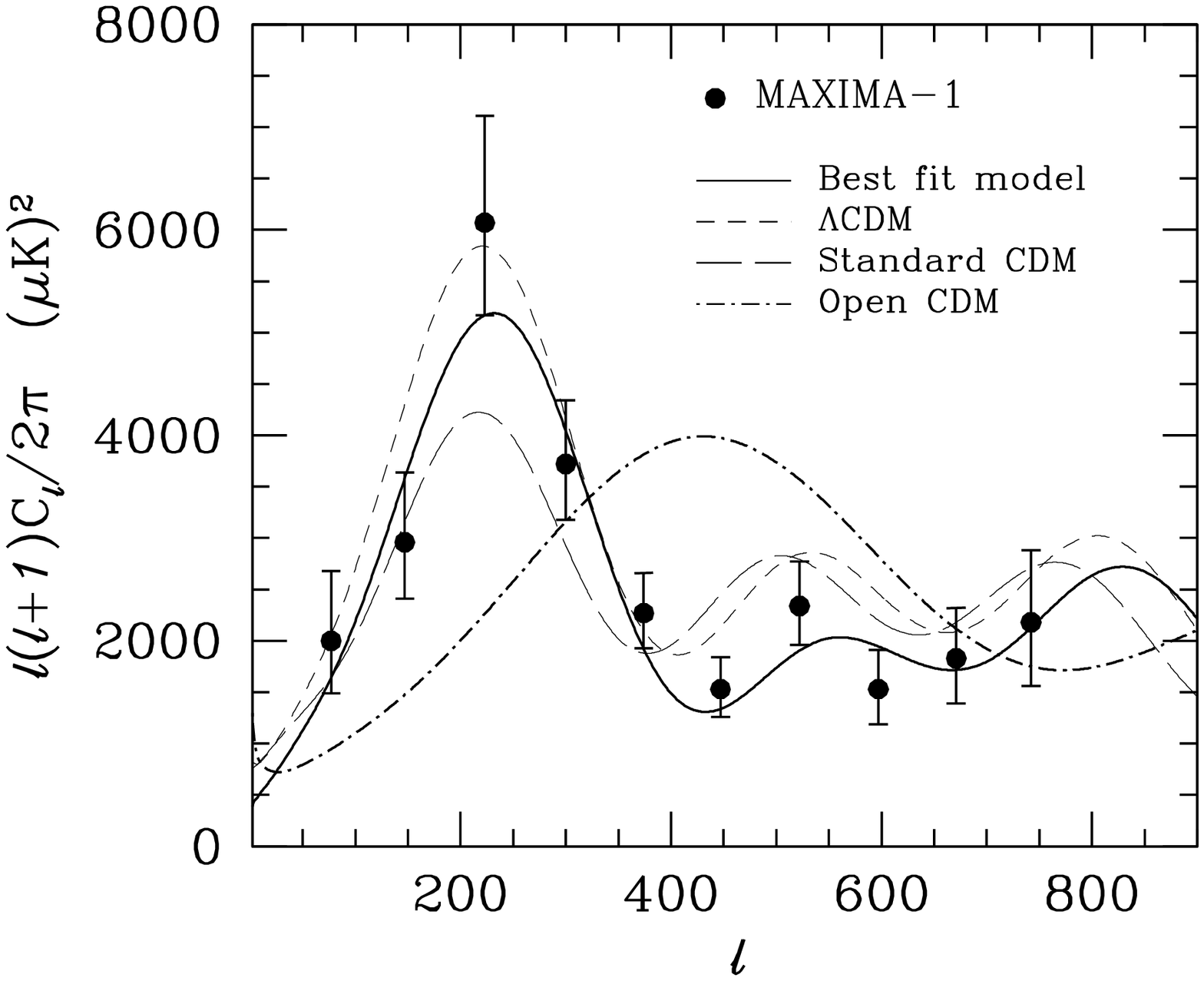}} 
{ \vskip -1cm\small
  F{\scriptsize IG}.~4.--- Angular power spectrum from MAXIMA-1 
  (Hanany et al. 2000). The solid line is the best fit model of our
  analysis, having ($\Omega_{\rm b},\Omega_{\rm
    cdm},\Omega_\Lambda,n_s,h) =
  (0.1,0.6,0.3,1.08,0.53$).  The long-dashed line
  is the Standard CDM model
  (0.05,0.95,0,1,0.5).  The short-dashed line
  is the $\Lambda$-CDM model with
  (0.05,0.35,0.6,1,0.65).  The dot-dashed line
  is the Open CDM model with $\Omega=0.3$.
\label{fig:fig5}}
\vskip 0.7cm

Finally, the agreement of our results with those obtained from other
observations lends further credence to the possibility that we are
considering the correct family of cosmological models.  First, we find
an independent constraint on the physical baryon density of the
universe: $0.02<\Omega_{\rm b}h^2<0.04$, which, although favoring higher
values than the BBN estimate of Burles et al. (1999), agrees with it 
within 2$\sigma$.
Second, we are able to constrain the amount of cold
dark matter in the universe: $\Omega_{\rm cdm}h^2=0.2^{+0.2}_{-0.1}$.  If
we combine this value with our constraint on $\Omega_{\rm b}h^2$ and
the MAXIMA-1/COBE/SN~Ia best fit value of $\Omega_{\rm m}\simeq 0.375$
derived from Figure~3, we find that the ``shape'' parameter (commonly
used in analysis of large scale structure; 
see e.g., Sugiyama 1995) is $\Gamma\equiv\Omega_{\rm
  m} h~\exp{(-\Omega_{\rm b} - 
\Omega_{\rm b} / \Omega_{\rm m})}=0.24 \pm 0.1$.
Together with our best estimate of $n_S = 1.08 \pm 0.1$,
this is consistent with the value $\Gamma=0.23 - 0.28 (1 - 1/n_S)$ which
emerges from a completely independent analysis of galaxy catalogs 
(Viana \& Liddle 1999; Liddle et al.\ 1995; Peacock \& Dodds 1994).

\acknowledgments We thank S. Jha and the High-Z Supernova Search Team
for kindly providing the combined likelihood from supernova
measurements used in this paper. PGF acknowledges fruitful discussions 
with A. Melchiorri. MAXIMA is supported by NASA Grants
NAG5-3941, NAG5-4454, by the NSF through the Center for Particle
Astrophysics at UC Berkeley, NSF cooperative agreement AST-9120005.
The data analysis used resources of the National Energy Research
Scientific Computing center which is supported by the Office of
Science of the U.S.  Department of Energy under contract no.
DE-AC03-76SF00098.  PA acknowledges support from PPARC rolling grant, UK.
PGF acknowledges support from the RS. AHJ and JHPW
acknowledge support from NASA LTSA Grant no. NAG5-6552 and NSF KDI
Grant no. 9872979. BR and CDW acknowledge support from NASA GSRP
Grants no. S00-GSRP-032 and S00-GSRP-031. We acknowledge use of CAMB.

\newpage

\centerline{ERRATUM} 

\vskip 1cm 

In the Letter ``Constraints on Cosmological Parameters from MAXIMA-1''
by A.~Balbi et al. (ApJ 545, L1 [2001]), an error was found in the
results: while it is claimed in the Letter that the optical depth of
reionization, $\tau_c$, took values bewteen $0$ and $0.5$, an error in
the computer code led to the results being based on $\tau_c=0$. We
have regenerated the database of models with the correct (published)
values of $\tau_c$, namely, $\tau_c$ = 0, 0.025, 0.05, 0.075, 0.1,
0.15, 0.2, 0.3, 0.5. The best-fit model quoted in the Letter remains
unchanged. However, we have found that the $95\%$ confidence level
constraints on the total energy density of the universe, the physical
density of baryons and the scalar spectral index of primordial
fluctuations change to: $\Omega=1.0^{+0.25}_{-0.30}$, $\Omega_{\rm
  b}h^2=0.030_{-0.010}^{+0.018}$ and $n_s=1.08_{-0.08}^{+0.32}$,
respectively.  The likelihoods for the remaining parameters are
effectively unchanged. In Figures 1 and 2, we present the figures
which should replace Figures 1 and 2 of the original Letter.
 
\vskip 1cm
 
\begin{multicols}{2}
  \vbox{\epsfxsize=8.5cm\centerline{\epsfbox{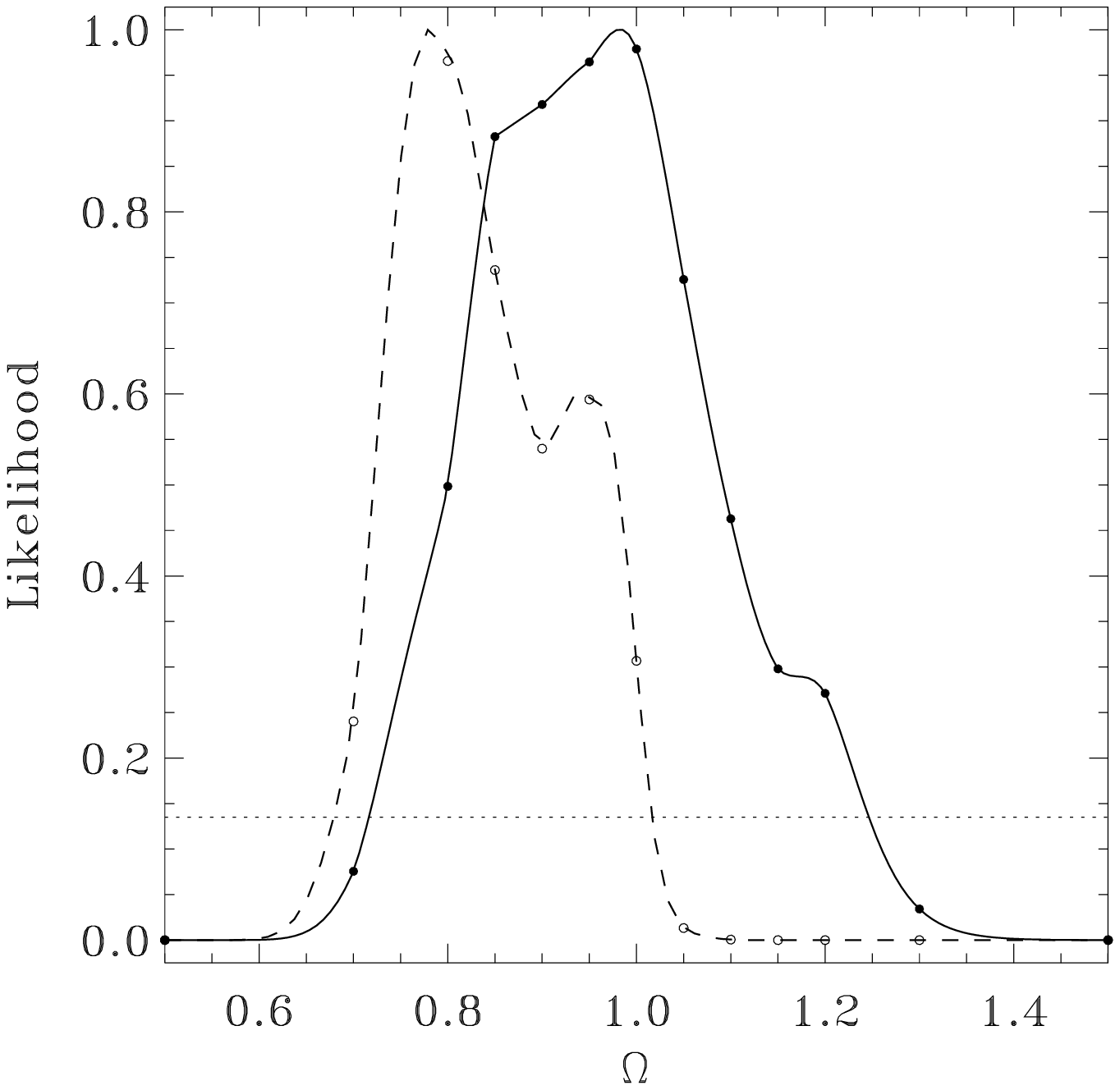}}} \vskip 0.3cm
  {\small F{\scriptsize IG}.~1.--- Likelihood function of the total
    energy density of the universe $\Omega$.  The solid line was
    obtained by maximizing over all other parameters over the ranges
    described in the text, while the dashed line was obtained by
    constraining $\Omega_{\rm b}h^2=0.0190\pm 0.0024$ and $H_0=65\pm
    7$~km~s$^{-1}$~Mpc$^{-1}$.  The intersections with the horizontal
    line give the bounds for $95\%$ confidence.}
     
  \vbox{\epsfxsize=8.5cm\centerline{\epsfbox{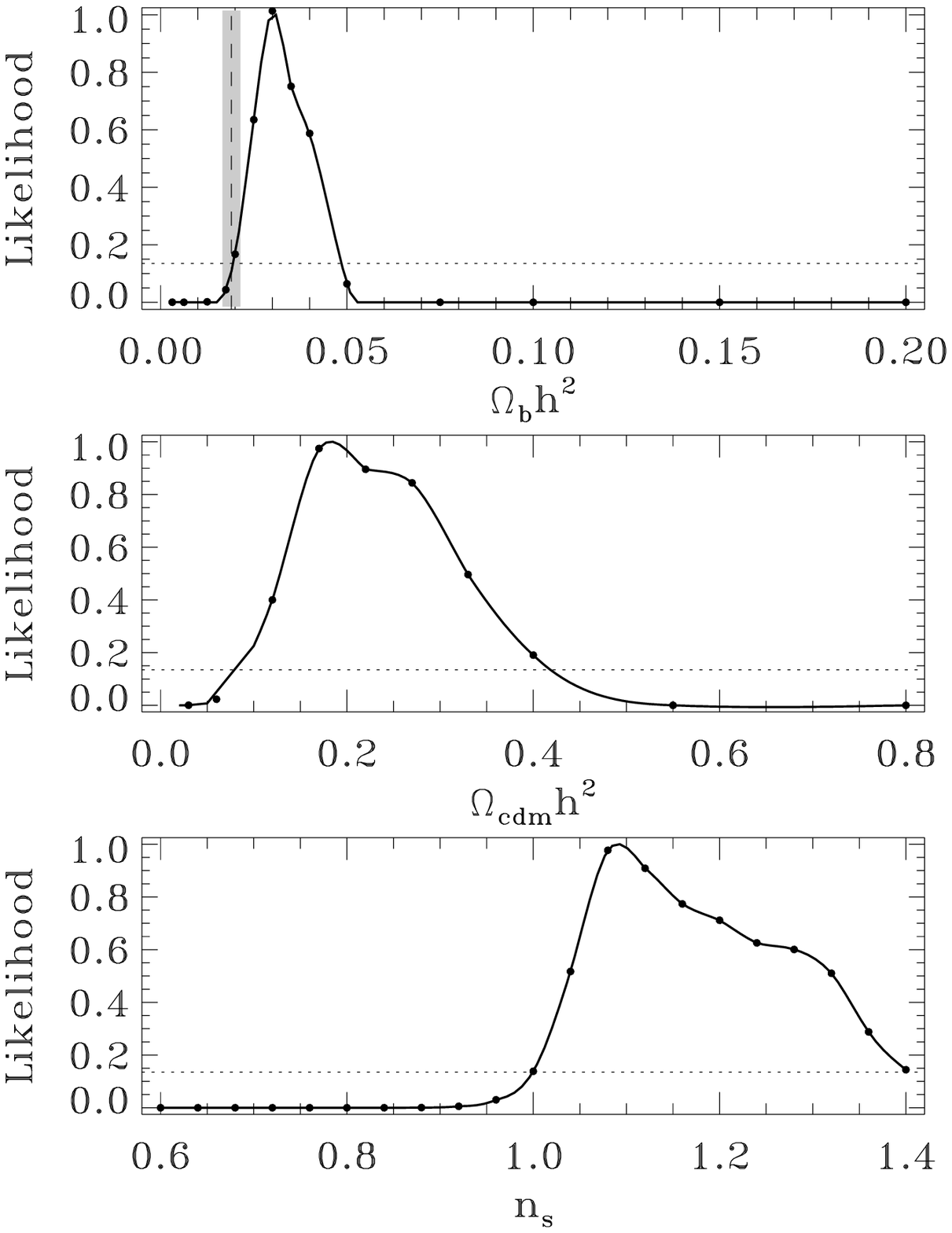}}} \vskip 0.5cm
  {\small F{\scriptsize IG}.~2.--- Likelihoods of three cosmological
    parameters: from top to bottom, the physical baryon density,
    $\Omega_{\rm b}h^2$, the physical cold dark matter density,
    $\Omega_{\rm cdm}h^2$, and the scalar spectral index, $n_s$.  For
    each of these parameters, the likelihood was maximized over all
    the remaining parameters.  The vertical band in the top panel
    represents the BBN constraint $\Omega_{\rm b}h^2=0.0190\pm 0.0024$
    (Burles et al. 1999). The intersections with the horizontal line
    give the bounds for $95\%$ confidence.}

\end{multicols}


\begin{references}
  
  \reference{} Bond, J.R., Jaffe, A.H., \& Knox, L. 1998, \prd, 57,
  2117
  
  \reference{} Bond, J.R., Jaffe, A.H., \& Knox, L. 2000, \apj, 533, 19
  
  \reference{} Borrill, J. 1999, to appear in Proceedings of the 5th
  European SGI/Cray MPP Workshop, Bologna, Italy, astro-ph/9911389
  
  \reference{} Burles, S., Nollett, K.M., Truran, J. N., \& Turner, 
  M.S.  1999, \prl, 82, 4176

  \reference{} Dodelson, S., Gates, E., \& Stebbins, A. 1996, \apj,
  647, 20
  
  \reference{} Dodelson, S., \& Knox, L. 1999, \prl, submitted,
  astro-ph/9909454
  
  \reference{} Doroshkevich, A.G, Zel'dovich, Ya., \& Sunyaev, R.
  1978, Sov. Astron., 22, 523
  
  \reference{} Efstathiou, G., \& Bond, J.R. 1999, \mnras, 304, 75
  
  \reference{} Freedman, W.L. 1999, astro-ph/9909076
  
  \reference{} G\'orski, K.M., et al. 1996, \apj, 464, L11
  
  \reference{} Hanany, S., et al. 2000, \apj, submitted
  
  \reference{} Kamionkowski, M., \& Kosowsky, A. 1999, Annu. Rev.
  Nucl. Part. Sci., 49, 77

  \reference{} Krauss, L.M., \& Turner, M.S. 1995, Gen. Rel. Grav., 27, 1137 

  \reference{} Lange, A.E., et al. 2000, \prd, submitted, astro-ph/0005004
  
  \reference{} Lewis, A., Challinor, A., \& Lasenby, A. 2000, \mnras,
  in press, astro-ph/9911177

  \reference{} Liddle, A.R., Lyth, D.H., Schaefer, R.K., Shafi, Q., \& Viana, 
  P.T.P.\ 1996, \mnras, 281, 531
  
  \reference{} Lineweaver, C.H. 1998, \apj, 505, L69
  
  \reference{} Knox, L., \& Page, L. 1999, \prl, submitted,
  astro-ph/0002162
  
  \reference{} Melchiorri A., et al. 1999, \apj, submitted,
  astro-ph/9911445
  
  \reference{} Ostriker, J. P. \& Steinhardt, P. J. 1995, \nat, 377,
  600
  
  \reference{} Peacock, J.A., \& Dodds, S.J. 1994, \mnras, 267, 1020
  
  \reference{} Perlmutter, S., et al. 1999, \apj, 517, 565
  
  \reference{} Riess, A.G., et al. 1998, \aj, 116, 1009
  
  \reference{} Seljak, U., \& Zaldarriaga, M. 1996, \apj, 469, 437
  
  \reference{} Stompor, R., \& Efstathiou, G. 1999, \mnras, 302, 735
  
  \reference{} Sugiyama, N. 1995, \apjs, 100, 281

  \reference{} Tegmark, M., \& Zaldarriaga, M. 2000, \apj, submitted,
  astro-ph/0002091

  \reference{} Tytler, D., et al. 2000, submitted to Physica Scripta,
  astro-ph/0001318

  \reference{} Viana, P.T.P., \& Liddle, A.R. 1999, \mnras, 303, 535 

\end{references}
\end{document}